\documentclass[fleqn,usenatbib]{mnras}

\usepackage[T1]{fontenc}

\DeclareRobustCommand{\VAN}[3]{#2}
\let\VANthebibliography\thebibliography
\def\thebibliography{\DeclareRobustCommand{\VAN}[3]{##3}\VANthebibliography}

\usepackage{graphicx}	
\usepackage{amsmath}	
\usepackage{amssymb}	
\usepackage{xcolor}
\usepackage[ separate-uncertainty = true, multi-part-units = repeat ]{siunitx}

\usepackage{tikz}
\usetikzlibrary{positioning,calc}

\usepackage{newtxtext,newtxmath}

\newcommand{\spar}{S_{\parallel}}
\newcommand{\sper}{S_{\bot}}
\newcommand{\vspar}{\vec{S}_{\parallel}}
\newcommand{\vsper}{\vec{S}_{\bot}}

\title[Alignment around voids]{Spin alignment around TNG300-1 voids}

\author[Dávila-Kurbán et al.]{
Federico Dávila-Kurbán,$^{1,2,3,4}$\thanks{E-mail: fdavilakurban@unc.edu.ar}
Marcelo Lares,$^{1,2,3}$
Diego Garcia Lambas,$^{1,2,3}$
\\
$^{1}$Instituto de Astronom\'ia Te\'orica y Experimental (IATE,
        CONICET/UNC), Córdoba, Argentina\\
$^{2}$ Observatorio Astron\'omico C\'ordoba, Argentina\\
$^{3}$ Consejo de Investigaciones Cient\'ificas y T\'ecnicas (CONICET),
Argentina \\
$^{4}$ Facultad de Matem\'atica, Astronom\'ia y F\'isica, Universidad
Nacional de C\'ordoba, Argentina
}

\date{Accepted XXX. Received YYY; in original form ZZZ}

\pubyear{2015}

\begin{document}
\label{firstpage}
\pagerange{\pageref{firstpage}--\pageref{lastpage}}
\maketitle

\begin{abstract}
Using a new statistical approach we study the alignment signal of galactic spins with respect to the center of voids identified in the TNG300--1 simulation.
We explore this signal in different samples of galaxies, varying their distance from the void center, mass, spin norm, local density, and velocity. We find a strong tendency (>9$\sigma$) of massive, high--spin, and low radial velocity galaxies to be aligned perpendicularly to the void--centric direction in a wide range of distances corresponding to 0.9 to 1.4 void radii. Furthermore, we find that in these subdense environments, local density is irrelevant in the amplitude of spin alignment, while the largest impact is associated to the galaxy void--centric radial velocity in the sense that those at the lowest expansion rate are more strongly aligned perpendicularly to the center of the void.  
Our results suggest that further analysis at understanding intrinsic alignments and their relation to large scale structures may probe key for weak lensing studies in upcoming large surveys such as Euclid and LSST.
\end{abstract}




\section{Introduction}

\begin{figure*}
  \centering
  \includegraphics[width=0.75\textwidth]{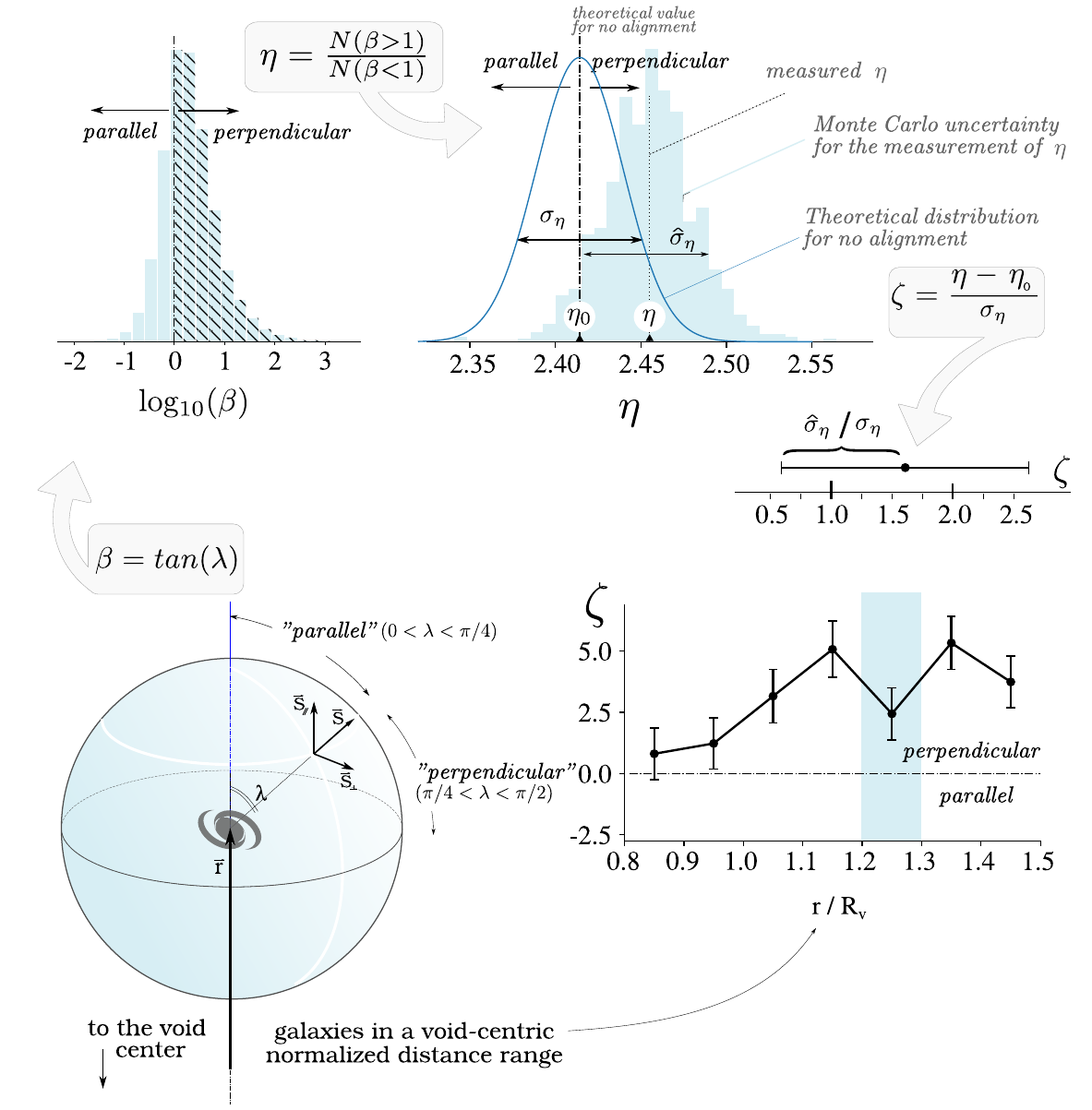}
  \caption{Scheme of the analysis performed in the search for alignment signals of galaxy discs around voids. We start with a $\lambda$ parameter, the acute angle between the spin vector and the galaxy vector position w.r.t. the void center. Then the ratio of galaxies with $\tan(\lambda)>1$ and with $\tan(\lambda)<1$ is compared with a theoretical distribution for a randomly oriented galaxy sample for different bins in radial distance to the void center. Finally, a normalized parameter is defined, $\zeta$, to show both the trend of alignment, if any, and its statistical significance; e.g. if $\zeta$>3, then this population of galaxies shows a trend of being perpendicularly aligned with a confidence of over 3$\sigma$.
  }
  \label{fig:abs}
\end{figure*}

Studies on the galaxy distribution in increasingly large spectroscopic surveys, for instance the Sloan Digital Sky Survey (SDSS; \citealt{York2000}), have shed light on the nature of a complex hierarchical network of structures, usually referred to as the cosmic web, composed of clusters, filaments, sheets, and voids (e.g., \citealt{Bond1996}). Preferential orientations, or \textit{alignments}, between galaxies, their underlying matter structures, and the aforementioned cosmic web are crucial aspects to further a more comprehensive understanding of gravity, the nature of matter, and structure formation in the Universe.

For a sufficiently large sample of galaxies in a homogeneous and isotropic universe one might expect galactic properties such as orientations and ellipticities to be random. For this reason, any detected net preferred orientation with regard to a given direction, any non-vanishing correlation between galaxy alignments, or any other phenomenon that indicates a local violation of isotropy, is usually linked to tidal gravitational forces acting on the galaxies at different evolutionary stages \citep{Peebles1969, Doroshkevich1970, White1984}. Furthermore, models of lensing effects, which explain coherent apparent distortions in galaxy images and help constrain cosmological models are themselves restricted by how well we understand any other possible sources of underlying coherent alignment \citep[e.g.][]{CroftMetzler2000, HeavensRefregierHeymans2000, HirataSeljak2004, Codis2015}.  

This work studies the orientation of galactic spins in void shells with respect to the void centric direction, within a simulation. 
%
%
Observational studies of orientations and alignments around low density environments such as voids are scarce, in part because, by the very definition of voids, the sample data to analyse are usually very small. However, there have been three widely discussed observational works (\citealt{Trujillo2006}, hereafter T06; \citealt{Slosar2009}, SW09; \citealt{Varela2012}, V12) that studied the orientation of galaxies around voids. What these works have in common is the use of the same void finder by \cite{Patiri2006}, which searches for the largest non-overlapping spheres within the survey volume devoid of galaxies above a certain threshold of brightness. They worked with SDSS data releases 3, 6, and 7, respectively. T06 additionally considered data from the 2dF Galaxy Redshift Survey and defined similar rest-frame magnitude thresholds. On the other hand, there were significant differences in the selection of the galaxy samples, and the measurement methods for their spins. T06 limited themselves to only selecting edge-on and face-on disc galaxies, while V12 fitted a thick-disc model to all galaxies that were classified as spirals by GalaxyZoo (\citealt{Lintott2008}).

The standard picture of tidal torque theory (\citealt{Lee2000, Lee2001, Lee2007}) postulates a preferential net alignment of the spin vector with the intermediate principal axis of the tidal shear tensor which lies tangentially to the surface of the void.
In agreement with this picture, T06 found a 99.7 per cent confidence level that spiral galaxies located on the shells of the
largest 
(>10~Mpc~$h^{-1}$)
cosmic voids have rotation axes that lie preferentially on 
the void surface.
SW09 found no statistical evidence for departure from random orientations; they argue that the results of T06 might possibly be a statistical fluctuation given that the catalogue used in SW09 is considerably larger and has a much better filling factor that dramatically increases the number of voids.
On the other hand, V12 considered voids with minimum radii of 15~Mpc~$h^{-1}$ and found a significant signal (>98.8 per cent) for the alignment of the spin of galaxies around these voids to be preferentially parallel to the radius vector, while for smaller voids this tendency disappears and the results are consistent with no special alignment.
Moreover, V12 also finds that the strength of the alignment depends on the distance between the galaxies and the void surface and that, regardless of void size, for galaxies farther than $\simeq$5~Mpc~$h^{-1}$ there is no preferential direction in the distribution of the alignments.
Regarding the disagreement with net tangential orientation (T06) or no orientation (SW09), V12 argue that the small size of the galaxy sample around voids with R~$\geq$~10~Mpc~$h^{-1}$ used in these works could mask the alignment signal that they find.

In recent years, as increasingly higher resolution simulations become available, there have several studies on alignments of spins as well as galaxy/halo shapes with respect to the various substructures of the cosmic web. Although it might be tempting to think of void shells, which are the focus of this work, as being equivalent to the ``sheet'' substructures of the cosmic web, it should be noted that the practical algorithms to identify them are significantly different (see e.g. the review \citealt{Joachimi2015}).

On the observational aspect, the scenario for spin alignments with sheets remains unclear. Using observations based on photographic plate data, \cite{Lee2002} and \cite{Lee2007} concluded that galaxy spins tend to lie within sheets, whereas using SDSS data, \cite{Tempel2013} and \cite{Zhang2015} found that galaxy angular momenta points preferentially perpendicular to the plane of the sheet, albeit with a weak signal in both types of alignment.
The latter results seem consistent with the void result of V12, however, simulation--based results generally coincide in finding that angular momenta lay preferentially parallel to planar structures (e.g. \citealt{Libeskind2013}), and this tendency seems to get stronger with more massive haloes (e.g. \citealt{Forero-Romero2014}). 
Studies of correlations of galaxy shapes located approximately around one void radius have been inconclusive due to the small galaxy sample \citep[e.g.][]{Reischke2019}. Despite this, \citet{dAssignies2022} recently suggested the existence of two regimes for alignment: a large-scale regime of over twice the void radius where the alignment would be radial and an intermediate regime with a projected distance up to 1.5 void radius where the shapes would be aligned tangentially. The authors find this latter regime to be interesting because it would depend directly on the mass distribution inside the void and is therefore useful to extract physical information and constraints. They confirm tangential alignment of shapes in this intermediate regime only in voids with radii between 30-40~Mpc~$h^{-1}$. A comparison with our work is difficult, however, since they employ projected distances and a different void identifier.



On the other hand, \citet{Codis2018} and \citet{Kraljic2019}, using the public \textsc{DisPerSE}\footnote{\url{http://www.iap.fr/users/sousbie/disperse.html}} (\citealt{Sousbie2013DisPerSE:Extractor}) algorithm to identify substructures in the Horizon-AGN and SIMBA simulations respectively, find a mass--dependent ``spin-flip'' for galaxies and haloes. Their results agree on the spin of low-mass galaxies being more likely to lie within the plane of sheets while massive galaxies preferentially having a spin perpendicular to the sheets. 

There are several reasons for the inconclusive results mentioned previously. From the observational point of view, a main difficulty remains in relating observed shapes with spin directions or ellipsoidal orientations, using different methods and possibly yielding different results with the same sample. The signal itself, additionally, seems to depend significantly on the parameters used to select subsamples of galaxies (such as luminosity, morphology, colours, etc), which obstructs a clear comparison between different works.
Furthermore, the small number of galaxies in voids imply that observational results in these environments have large statistical uncertainties (see \citealt{Zhang2015}). 
For these reasons, in addition to the improving computing power, simulation--based studies have thrived in this area, particularly in the search for alignments of baryonic and dark matter haloes with regards to the components of the large scale structure (see e.g. \citealt{Codis2018} and \citealt{Kraljic2019}). 

In this work we study alignment of galactic spins in void shells with respect to the void center using a well--established void identifier (\citealt{Ruiz2015}) and analyze the dependence of the strength of the signal with mass, spin norm, velocity, and local density. By including velocities in our spin alignment analysis we study a dynamical aspect that has not been sufficiently explored before. Finally, we employ a novel method that uses robust and well--behaved statistical parameters to reject or accept the null--hypothesis of no alignment (Dávila Kurbán et al., submitted). Hopefully this new approach will provide a useful perspective on the issue of galaxy orientations.

The outline of our work is as follows.
Sec.~\ref{sec:method} will present a brief summary of the parameters used to explore, describe, and quantify the alignment signal and how they vary with spin norm, mass, velocity and local density.
We describe the simulation, the void identification algorithm, and the population of voids and galaxies in Sec.~\ref{sec:data}.
The main results are presented and discussed in sections \ref{sec:results} and \ref{sec:conclusions}.

\section{Method and statistics}
\label{sec:method}

The spherical symmetry of voids, both in their geometry and dynamics, allows for a specific direction in which to analyse galactic orientations: the radial direction. Given the problem of vector orientations around a central point we will define the parameters $\beta$, $\eta$, and $\zeta$ that will allow us to study the orientation of galaxies and detect possible excesses with respect to a random distribution. These three parameters are formally introduced and analysed in Dávila Kurbán et al. (submitted), however, the basic definitions are outlined below. 
The motivation behind this approach is to develop a robust statistic for the measuring of alignment signal that does not rely on Monte Carlo simulations for an estimation of its statistical significance (Dávila Kurbán et al., submitted).

Additionally, complementing the description of this section, Fig.~\ref{fig:abs} shows a schematic summary for the reader as a quick refresher of the definitions of the parameters, how they relate to one another, and, ultimately, how we start from the measurement of an angle to the visual representation of alignment signal we use to show our results.

\subsection{Definition of the $\beta$ parameter}
\label{sec:betadef}

Given the radial direction $\hat{r}$ of unit norm, one can calculate the parallel and perpendicular components of the spin vector $\vec{S}$:

$$
S_{\parallel} = |\vec{S}|\;|\cos(\lambda)| = |\vec{S} \cdot \hat{r}|, \qquad \textrm{and} \qquad S_\perp = \sqrt{ \vec{S}^2 - S^2_{\parallel} },
$$
 
\noindent
where $\vsper$ is the perpendicular component of the radial direction, $\hat{r}$, and $\vspar$ is the parallel component of said direction, so that $\vec{S}=\vsper+\vspar$. By taking the absolute value of $\mathrm{cos}(\lambda)$ we determine that $\lambda$ is in fact the acute angle between the radial direction $\hat{r}$ and the spin vector $\vec{S}$.

The distribution of the acute angle $\lambda$ can be used to analyse alignments of the spin vectors, and given the relation of this to the components of the vector, the latter can be used to determine the orientations. Therefore, we define:

\begin{equation}
    \beta = \frac{\sper}{\spar} = \tan(\lambda).
    \label{eq:beta}
\end{equation}

Now $\beta$ is also a measure of the orientation of the spin vector $\vec{S}$. Note that given our definitions of $\vsper$ and $\vspar$, our parameter $\beta$ is always positive: 

$$
0 \le \lambda \le \pi/2; \qquad 0 \le \beta \le \infty
$$

Spin vectors with $\beta>1$ lay preferentially on the \textit{perpendicular} direction with $\pi/4 < \lambda < \pi/2$, while those with $\beta<1$ have a preferential orientation on the \textit{parallel} direction with $0 < \lambda < \pi/4$.

Given that the probability distribution function of $\beta$ is pathological (Dávila-Kurbán et al., submitted), we cannot use this parameter directly if we want to develop a statistical method that is robust. Instead we use $\beta$ to define below the parameters $\eta$ and $\zeta$.

\subsection{Definition of the $\mathrm{\eta}$ and $\zeta$ parameters}
\label{sec:etadef}

Given a population of spin vectors with a measured $\beta$ parameter, we need an robust estimator to analyse the statistical tendency in said population of preferring a perpendicular or parallel direction, and measure whether this tendency is sufficiently different from random behaviour.

We consider the fraction of values of $\beta$ that are greater than some critical value. Given that when the perpendicular and parallel components are equal there is no preference for either direction, we propose that the critical value be $\beta=1$. Therefore, we define the parameter

\begin{equation}
    \eta = \frac{n(\beta>1)}{n(\beta<1)}
    \label{eq:eta}
\end{equation}

\noindent
where n is the number of observations of a sample that fulfills the conditions indicated in parentheses. 

It can be shown (Dávila-Kurbán et al., submitted) that this variable $\eta$, under the null hypothesis of no alignment, closely follows a Gaussian distribution and is therefore completely described by its first two moments, given by:

\begin{equation}
    E(\eta) \equiv \eta_0 = \frac{1}{\sqrt{2}-1} \simeq 2.4142
\end{equation}

\begin{align}
    Var(\eta) &= \left[\left(\frac{1}{Nq}\right)^2 + \left(\frac{p}{Nq^2}\right)^2\right] Npq + 2\frac{1}{Nq} \frac{p}{Nq^2} Npq \nonumber \\
    &\simeq \frac{28.1421}{N},
\label{eq_var_eta}
\end{align}

\noindent
where $p=1/\sqrt{2}$, $q=1-p$, and N is the total size of the sample. 

\begin{figure}
    \centering
    \includegraphics[width=1\columnwidth]{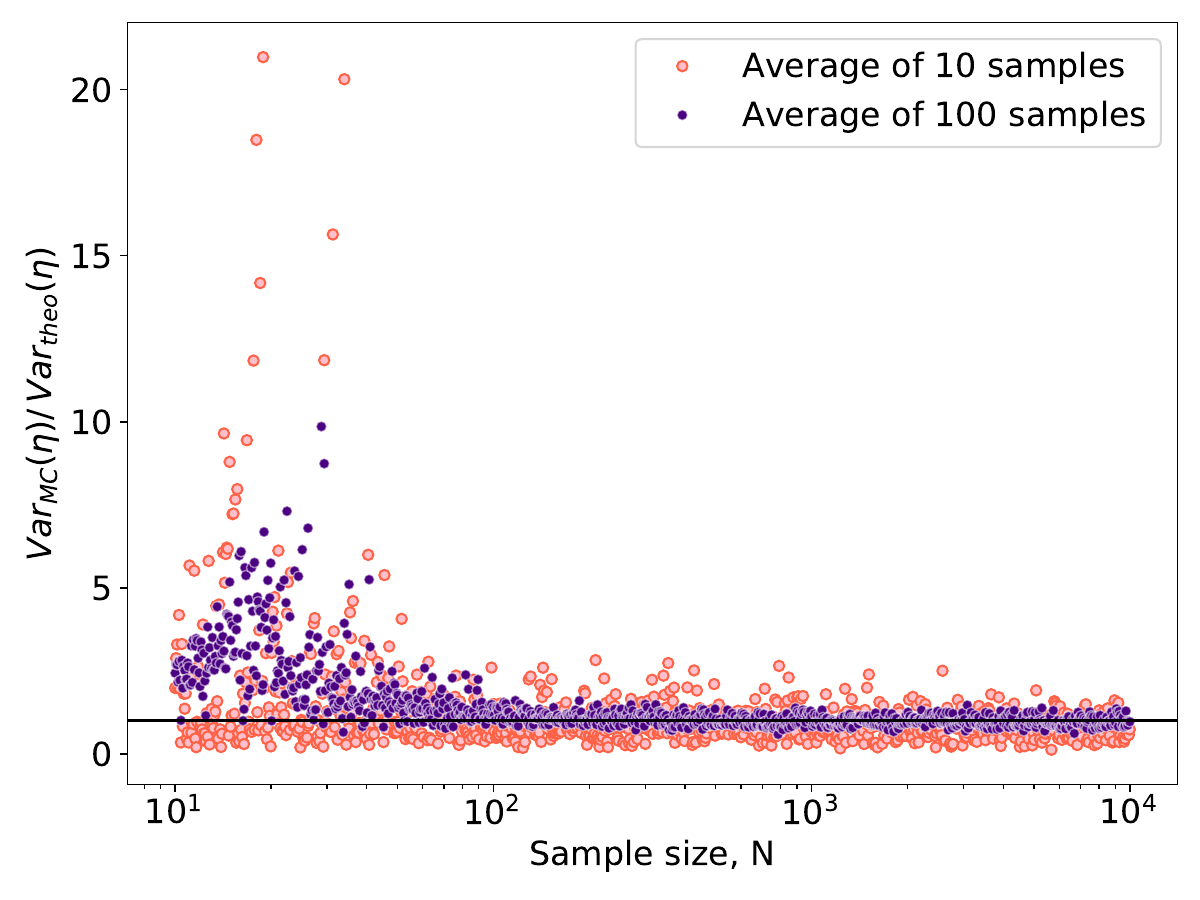}
     \caption{Ratio of the Monte Carlo estimation of the variance of the alignment parameter $\eta$, $\mathrm{Var_{MC}}$, and its derived theoretical expression, $\mathrm{Var_{theo}}$ (Eq.~\ref{eq_var_eta}). For sample sizes greater than approximately one hundred, the theoretical expression matches the Monte Carlo estimation.}
    \label{fig:eta_variance}
\end{figure}

The expected value, $E[\eta]$, can be understood as the volume ratio of the two sections within a sphere delimited by a 45 degree angle from a reference direction. A statistical formalism for the derivation of this value is detailed in Dávila-Kurbán et al. (submitted), as well as the derivation of the variance (Eq.~\ref{eq_var_eta}).
%
Here, in Fig.~\ref{fig:eta_variance} we show the ratio of Monte Carlo estimations of the variance of $\eta$ and the derived theoretical value, Eq.~\ref{eq_var_eta}, with different sample sizes. It can be observed that for sample sizes larger than roughly 100 the theoretical expression is a suitable estimation of the variance of $\eta$ (this criterion is met throughout this work, see Sec.~\ref{sec:data_gals}).

Finally, we define $\zeta$, the variable that will be used to express alignment signal, by normalizing $\eta$ as follows:

\begin{equation}
    \zeta \equiv \frac{\eta-\eta_0}{\sigma_{\eta}(N)}.
\end{equation}

\noindent
where $\sigma_{\eta}(N)$ is calculated with the square root of Eq.~\ref{eq_var_eta}. 
Note that $\zeta$ follows a normalized Gaussian distribution, and so $\zeta>0$ indicates a preferentially 
\textit{perpendicular} orientation, $\zeta<0$ indicates a preferentially \textit{parallel} orientation, and absolute values above 1, 2, and 3 indicate a confidence level of 1--, 2--, and 3--$\sigma$ respectively. An estimation for the error of $\zeta$ is calculated using a boostrap resampling technique, represented with error bars in Fig. \ref{fig:abs}.

\section{Data}
\label{sec:data}

\subsection{TNG300-1 Simulation}

We apply the previously described method to galaxy data from the IllustrisTNG project (TNG, \citealt{Pillepich2018b, Marinacci2018, Naiman2018, Springel2018, Nelson2018, Nelson2019a,  Nelson2019b, Pillepich2019}). IllustrisTNG is a suite of cosmological magneto-hydrodynamic simulations obtained with the moving--mesh code AREPO (\citealt{Springel2010}), and adopting the Planck cosmology (\citealt{Collaboration2016}): $\Omega_\mathrm{m}= 0.3089$, $\Omega_\mathrm{b}= 0.0486$, $\Omega_\mathrm{\Lambda}= 0.6911$, $\sigma_8= 0.8159$, $\mathrm{ns}= 0.9667$, and $\mathrm{h}= 0.6774$. These simulations present exhaustive models for galaxy formation physics, and improve upon their predecessor, Illustris, by including magnetic fields and improving galactic wind models and AGN feedback. The TNG project encompasses three different volumes with identical initial conditions and physical models: TNG50, TNG100, and TNG300. In particular, we employ the TNG300-1 with a periodic box of 205Mpc~$h^{-1}$, the largest box with highest resolution from the suite. The haloes (groups) and subhaloes (galaxies) in TNG are found with a standard friends-of-friends (FoF) algorithm with linking length $b=0.2$ (in units of the mean interparticle spacing) run on the dark matter particles, and the SUBFIND algorithm (\citealt{Springel2001}) respectively. The latter detects substructure within the groups and defines locally overdense, self-bound particle groups, where the baryonic component in the substructure is defined as a galaxy. We analyze the simulations at the final redshift, $\mathrm{z}=0$, considering galaxies with stellar mass of $10^{9}M_{\sun} \leq M_{\star} \leq 10^{13}M_{\sun}$.
The spin of the galaxies in the TNG suites is defined as the total spin per axis computed as the mass weighted sum of the relative coordinate times relative velocity of all member particles. The lower cut in mass mentioned above allows us to employ only galaxies in which the spin is well defined.

\subsection{Void identification and their galaxy population}
\label{sec:data_voids}

\begin{figure}
  \centering
  \includegraphics[width=0.5\textwidth]{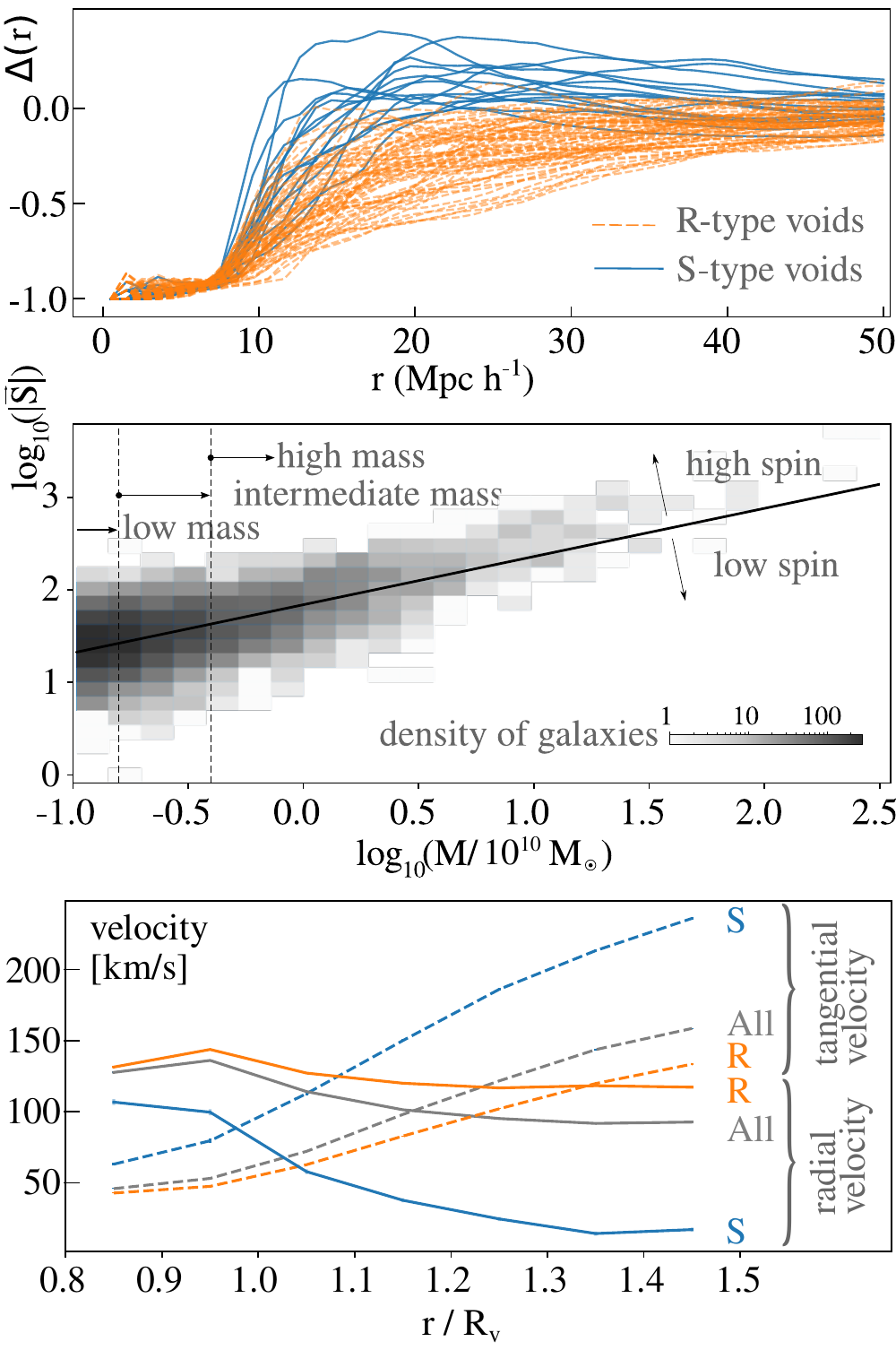}

  \caption{This figure encompasses several aspects of the data we work with. The upper panel shows the density profiles of the voids we identified in the simulation, and their classification into R--type and S--type voids. The middle panel plots the logarithm of the norm of the spin vector $\vec{S}$ of a galaxy as a function of the logarithm of its mass in units of $M_{\sun}$. The solid line is a linear regression, which we use to classify the galaxies into ``high'' and ``low'' spin, while the dotted vertical lines correspond to the logarithmic mass values: -0.8 and -0.4 which classify the galaxies into ``low'', ``intermediate'' and ``high'' mass. Finally, the bottom panel shows radial and transverse velocity as a function of distance to the center of the void in units of void radius. Transverse velocity increases with distance, as expected, given that the closer galaxies are to outer structures, they are more likely to be affected by non-radial gravitational pulls. Radial velocity, on the other hand, peaks before 1$\mathrm{R_v}$ and the behaviour differs, as expected, for R-- and S--type voids; the curve of the former averages at 120km~$\mathrm{s^{-1}}$ while that of the latter drops steadily. 
  Positive radial velocity indicates that the environments we are studying are very much in expansion, especially R--type voids, and must therefore be underdense.
  }
  \label{fig:data}
\end{figure}

The identification of voids in the simulation follows the algorithm described in \cite{Ruiz2015}, a modified version of previous algorithms presented in \cite{Padilla2005} and \cite{Ceccarelli2006}. The algorithm estimates the density profile with a Voronoi tessellation over density tracers, in particular, in this work, TNG galaxies. Underdense regions are obtained by selecting Voronoi cells below a density threshold and are selected as void candidates. Centered in these cells, the integrated density contrast $\Delta(r)$ is computed at increasing values of $r$. Void candidates are then selected as the largest spheres satisfying the condition $\Delta(\mathrm{R_{v}}) < 0.9$ where $\mathrm{R_{v}}$ is the void radius. Void centers are then randomly displaced so that the spheres are allowed to grow. This is done because the algorithm is likely to yield spherical voids where their shells do not precisely fit with the surrounding structures, and the recentering procedure provides structures with borders that better agree with the surrounding local density field. Finally, the void catalog comprises the largest underdense, non overlapping spheres of radius $\mathrm{R_{v}}$. 
After applying this algorithm to the TNG300-1, using subhaloes as tracers, and cutting off shot--noise voids, we are left with a sample of 82 voids with radii in the range 7-11Mpc~$h^{-1}$. 


Void surroundings can provide physical insight on the nature and evolution of void properties, since their hierarchy stems from the mass assembly in the growing structure nearby (\citealt{Sheth2004, Paranjape2012}). Some voids collapse onto themselves with their surrounding structure while other voids remain as underdense regions. These two types of evolution are determined by the surrounding density: voids surrounded by an environment resembling the mean background density will expand and remain an underdense region, known as a void--in--void type, whereas if the void is surrounded by an overdense shell, a void--in--cloud system, it will likely shrink under the collapse of the shell. These two evolutionary behaviours can be identified by calculating the cumulative radial density profiles (\citealt{Ceccarelli2013, Paz2013, Ruiz2015}), such as shown in the top panel of Fig. \ref{fig:data}. These profiles can be used to classify the void sample into void--in--void, called R--type (\textit{rising} type), and void--in--cloud systems, dubbed S--type (\textit{shell} type). Voids with a smoothly rising integrated radial profile are classified as R--type voids, while those embedded in a globally overdense region are classified as S--type.
We employ subhaloes as tracers of the density around voids; the same tracers employed to identify the voids.
The top panel in Fig.~\ref{fig:data} shows, in solid orange lines, the density contrast of shell voids (S--type) while the dashed blue lines represent those of rising voids (R--type). 
In practice, the classification of voids is done by evaluating $\Delta(3\mathrm{R_v})$, labeling them as R--type or S--type when the value is under or over zero, respectively.
We study the alignment signal in each of these two types of voids as well as in the complete void sample. In that analysis we will refer to the complete void sample as ``all void types''.

\subsection{Galaxy velocities and environment}
\label{sec:galaxyvel}

In this section we study the dynamics of these galaxies and their local environment in order to better interpret the results. To this end we explore the mean radial and transverse velocity of galaxies as a function of their radial distance from the void center (both R-- and S--type voids), as shown in the bottom panel of Fig. \ref{fig:data}. As expected, at approximately 1$\mathrm{R_v}$ and beyond, void--centric radial velocities starts to decrease while transverse velocities continue to rise driven by the more frequent overdense structures such as filaments and massive clusters. 

We stress the fact that in the range of void--centric distances analysed, the radial velocities of galaxies are still quite large, generally in the range 25 to 120km~s$^{-1}$ for S-- and R--type voids, respectively. This confirms that these regions are still in global expansion, associated to an underdense large--scale environment.

To confirm this, we computed the accumulated density contrast, same as in the top panel of Fig.~\ref{fig:data}, but as a function of the void radii in the 0.8-1.5$\mathrm{R_v}$ range of distances. 
The density contrast of R--type voids is below -0.25 for this entire distance range, with only a few voids actually surpassing the -0.50 value at the furthest distances of $\sim$1.5$\mathrm{R_v}$. More on this below.

\subsection{Properties of galaxies in voids and their classification}
\label{sec:data_gals}

We focus in this subsection at exploring the alignment dependence on different galaxy characteristics. We consider intrinsic properties, mass and total spin, and also their local density environment and expansion velocity with respect to void centre. In order to study the latter we consider increasingly larger, non--overlapping shells of width 0.1$\mathrm{R_v}$. Also, we divide the population within the shells into ``high'' and ``low'' velocity samples with respect to the median values. We take account of the local density density using the $\Sigma_5$ statistical parameter defined in Sec.~\ref{sec:res_s5}. We find that the expansion velocity and local density are independent variables so it is feasible to study them separately. 

In order to distinguish the galaxy populations into high and low spin systems requires to analyse the spin mass correlation.

The middle panel of Fig. \ref{fig:data} shows the spin and mass of a population of galaxies selected at random, corresponding to a shell of 1.0-1.1$\mathrm{R_v}$ of inner and outer radii of a void with $\mathrm{R_v}$~$\simeq$~8Mpc~$h^{-1}$. A simple differentiation into high/low spin galaxies is done by performing a linear regression on the spin-mass relation as shown with a solid line in the middle panel of Fig. \ref{fig:data} which divides the sample into similar number of objects. We have considered three mass ranges corresponding roughly to terciles of the sample: 
\mbox{$M_1=10^{9.2}M_\odot\,\text{and}\,M_2=10^{9.6}M_\odot$}; 
these two limits are shown in vertical dashed lines.

The total number of galaxies in shells of 0.8-1.5$\mathrm{R_v}$ around the 82 identified voids is $\mathrm{N_{All}}$~=~413864. The numbers in each of the seven radii bins, or equivalently, seven shells with a depth of 0.1$\mathrm{R_v}$ around R-- and S--type voids are, respectively: $\mathrm{N_R}$~=~[5666, 8855, 29273, 44152, 60364, 75188, 93348] and
$\mathrm{N_S}$~=~[1015, 1872, 6770, 12946, 18412, 25741, 30262].


\section{Results}
\label{sec:results}

\begin{figure}
    \centering
    \includegraphics[width=1\columnwidth]{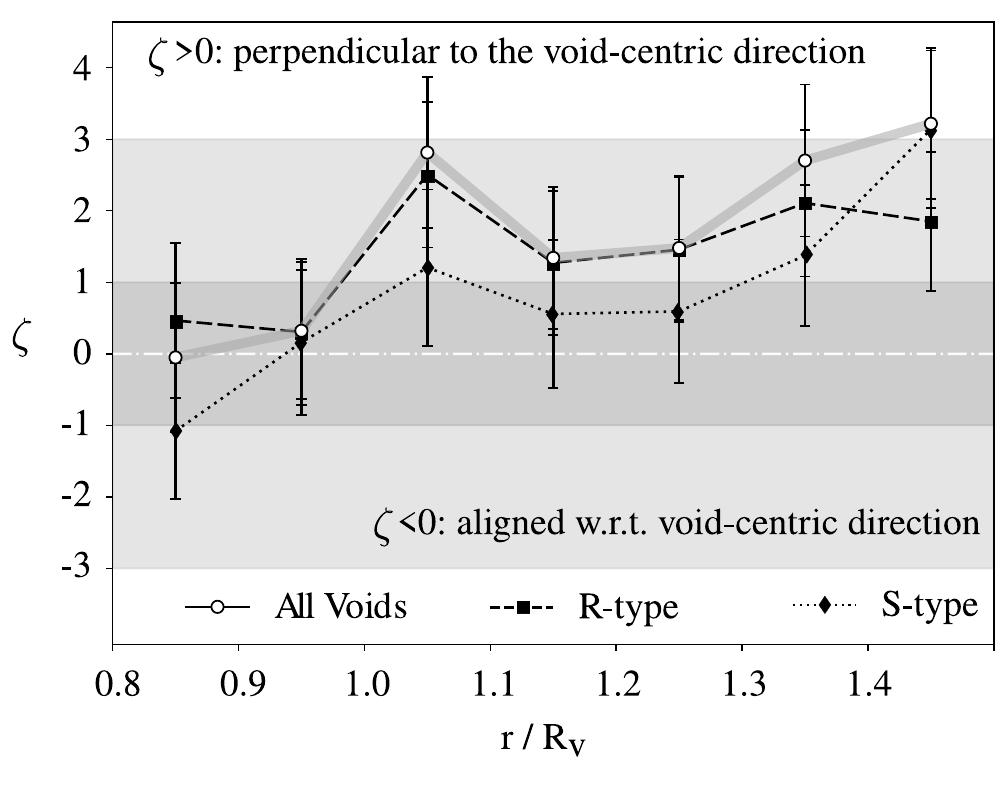}
    \caption{Alignment signal as a function of distance to the void center in units of void radius, $\mathrm{\zeta(r\, R_v^{-1}})$. The long and short dashed lines represent the signal around R-- and S--type voids respectively, while the solid line is represents the results for all void types. There is a noticeable peak of perpendicular signal, or $\mathrm{\zeta>0}$, in the shell centered in 1.05$\mathrm{R_v}$. Further away from this peak, the signal for S--Voids, while generally less than that of R--type voids, increases, surpassing the latter in the furthest bin of distance. Although nothing too significant, i.e. no signal above 3$\sigma$ of confidence, there is a general trend of perpendicular alignment for the entire, non--differentiated, galaxy sample.}
    \label{fig:eta_nosubsample}
\end{figure}

\begin{figure*}
\centering
\includegraphics[width=.9\textwidth]{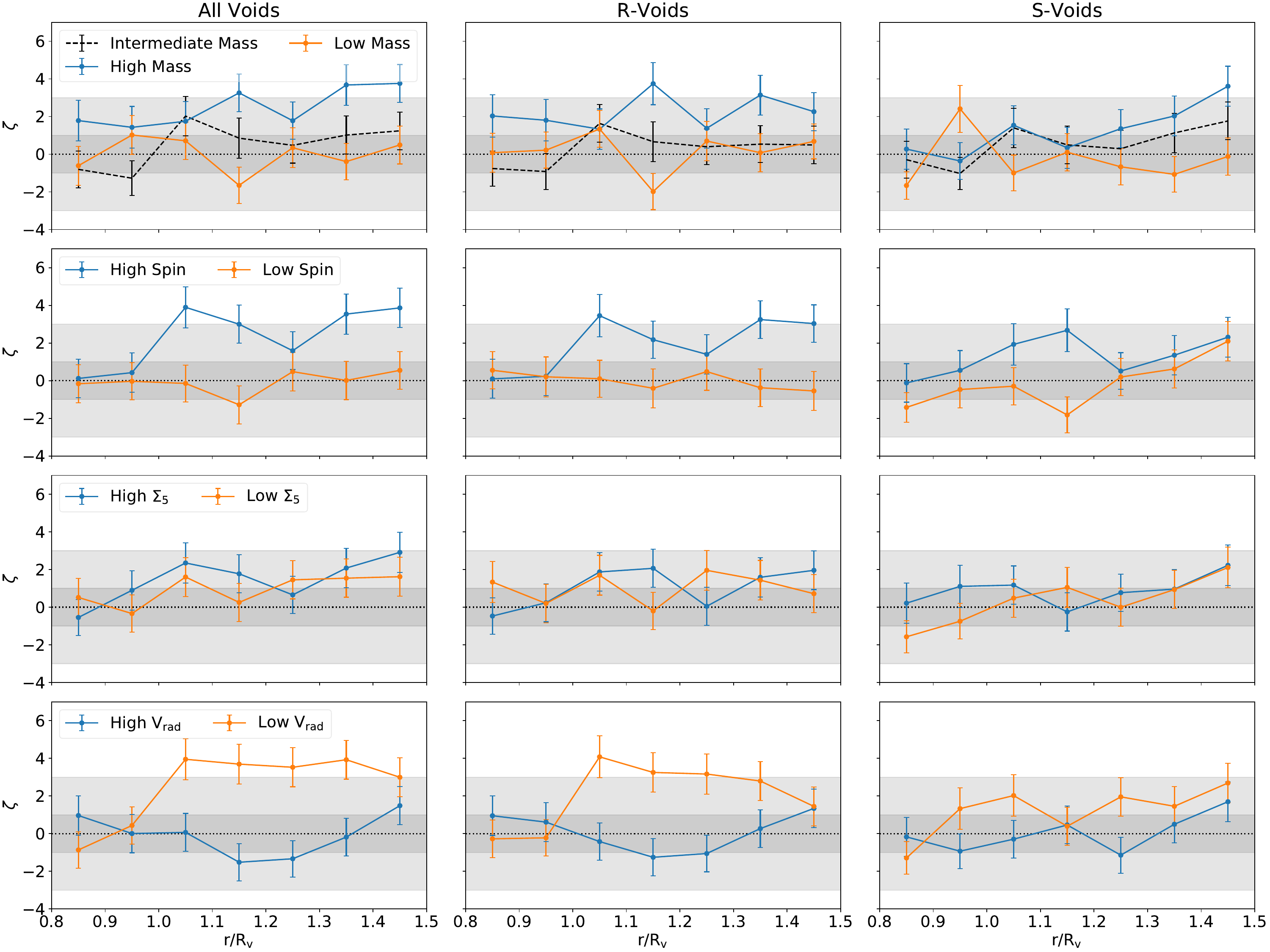}
\caption{
Alignment signal results as a function of radial distance to the void center,  $\mathrm{\zeta(r\, R_v^{-1}})$, for different galaxy samples. Blue lines represent samples with high values of the filtering galactic property while orange lines shows the samples with lower values. For completion we included the intermediate mass range in the first row indicated with a dotted black line. Strongly shaded regions encompass $1\sigma$ level of confidence while the light shade represents $3\sigma$ confidence regions. These regions correspond to uncertainties of reference samples and are calculated with the theoretical expression from the derivation of $\zeta$ (Eq. \ref{eq_var_eta}), while the error bars of the signal are obtained from bootstrap resampling.
Signals
of over $3\sigma$ are found when filtering for high mass, high spin, and low radial velocity. Filtering for high and low $\Sigma_5$ appears to have no significant effect. The galactic property that yields the strongest signal in this range of radial distances is low radial velocity.
}
\label{fig:eta_all}
\end{figure*}

In this section, we explore the alignment signal of galactic discs and its relation with galactic
properties and environmental features, such as the spin norm, the mass, the velocity, and the local density of galaxies.
To that end, we make use of the $\zeta$ parameter, defined in Sec.~\ref{sec:etadef}.
%
%
%
Our analysis aims at determining what properties of the galaxy sample produce a significant change in the alignment signal.
To study the dependency of the alignment signal $\zeta$ with respect to the radial position $r$ we first stack galaxies belonging to the full void sample, and then consider the rising (R--type) and the shell (S--type) void samples separately.
We select galaxies within spherical shells of $0.1\mathrm{R_v}$ depth,
%
from 0.8$\mathrm{R_v}$ to 1.5$\mathrm{R_v}$, comprising a total of 7 bins. 

Once we have determined the population of galaxies to be studied, we split it into ``low'' and ``high'' values of the galactic property we are studying, e.g. low and high spin, with the exception of mass which also has an ``intermediate'' classification, as discussed in Sec.~\ref{sec:data_gals}.
Then, for each of the subsamples we calculate $\zeta(\mathrm{r\, R_v^{-1}})$, i.e. the alignment as a function of the radial position with respect to the void center, and analyze if it changes significantly for the different subsamples.
With this approach, we look for parallel ($\zeta<0$) or perpendicular ($\zeta>0$) alignment trends.
First, we show the alignment signal with no galaxy classification, i.e. the signal of the entire galaxy population in the selected distances from the center of the void, (Fig.~\ref{fig:eta_nosubsample}). 
Then we present the results for each of the subsamples determined by the general classifications mentioned above (Fig.~\ref{fig:eta_all} and Secs.~\ref{sec:res_mass} through \ref{sec:res_vel}). 
Finally, in Sec.~\ref{sec:res_comb} we study the alignment in every possible combination of the mentioned selection criteria.

In Fig.~\ref{fig:eta_nosubsample} we show the alignment signal, quantified by the averaged values of the $\zeta$ parameter as a function of the
distance to the void center, considering the full sample of voids, along with the R--type and 
S--type subsamples.
The light and dark grey regions represent 3$\sigma$ and 1$\sigma$ significance, respectively, calculated using Eq.~\ref{eq_var_eta}, while the error bars are calculated with 1000 bootstrap resamplings of the galaxies in each distance bin across the stacked voids in consideration.
A general trend can be seen favouring a perpendicular alignment signal of the spin vectors w.r.t. the void--centric direction.
This result can be interpreted as suggesting a preference of discs 
to be found perpendicularly to the void center (i.e., $\zeta>0$).
The significance of this probability excess is between 1 and 3$\sigma$ for the full sample in distances between 1 and 1.5 void radii.
In spite of a marginal signal, it is consistently positive in a wide range, covering at least five bins in normalized distance.
%
%
In general, the signal for R--type voids seems to be stronger than that for S--type voids, and there is a noticeable alignment peak for both types in the shell centered in 1.05$\mathrm{R_v}$.
For further away bins, the signal for S--type voids increases faster than that of R--type voids. 
%

In Fig.~\ref{fig:eta_all} we show the results of splitting the galaxy sample according to high and low mass, spin, local density, and velocity. 
%
%
The blue and orange lines represent populations with high and low values of the parameter of interest, respectively.
We will explore the results of splitting the sample with respect to the mentioned parameters in the following subsections.

\subsection{Alignment dependence on mass}
\label{sec:res_mass}

We follow the mass classification discussed above and in Sec.~\ref{sec:data_gals}, and show in the first row of Fig.~\ref{fig:eta_all} the results of the alignment signal for samples with low, intermediate, and high mass.
Low and intermediate mass samples are mostly consistent with no alignment signals.
However, a consistent signal above 1$\sigma$ is found when analyzing high mass samples.
In particular, shells with significant perpendicular signal, i.e. above the 3$\sigma$ shaded region, are found centered in 1.15$\mathrm{R_v}$ around R--type voids, and 1.45$\mathrm{R_v}$ around S--type voids.
Taking all void types into account, a stronger signal is also found in shells centered in 1.35$\mathrm{R_v}$

These results do not indicate a clear mass--dependent spin flip in void shells with this mass binning.
However, the results suggest a trend of changing from no alignment to perpendicular alignment with increasing mass.
There is a phenomenon resembling a spin flip for shells centered in 1.15$\mathrm{R_v}$ around R--type voids, in which low-mass galaxies seem to show alignment signal of $\zeta\simeq$-2\textpm1, however this is not a strong enough signal for us to conclude in favor of the existence of a spin flip in this instance.

\subsection{Alignment dependence on spin}
\label{sec:res_spin}

\noindent

We follow the spin classification discussed above and in Sec.~\ref{sec:data_gals}, and show in the second row of Fig.~\ref{fig:eta_all} the results for the alignment signal for low and high spin samples. 

Low spin galaxies results are mostly consistent with no alignment across all void classifications, i.e. $\mathrm{\zeta(r\, R_v^{-1}})\simeq0$. On the other hand, high spin galaxies, i.e. galaxies that have acquired relevant rotation, show a strong and significant ($\zeta\geq3$) tendency to be perpendicular for $\mathrm{r>1R_v}$, especially in R--type voids. Shells that exhibit an above--3$\sigma$ signal are centered around 1.05, 1.15, 1.35 and 1.45$\mathrm{R_v}$ in the all void types sample, and 1.05$\mathrm{R_v}$ in R--type voids. No such signal is found in S--type voids. Furthermore, for S--type voids, there seems to be no statistically significant difference in alignment between low and high spin galaxies for $\mathrm{r>1.2R_v}$.


\subsection{Alignment dependence on local environment density}
\label{sec:res_s5}

The nearest neighbour approach studies the environment density by considering a variable scale estimator. Usually the surface density parameter is calculated as $\Sigma_n = n /\pi r^{2}_n$, where n is the number of neighbours within a circumference with radius equal to $\mathrm{r_n}$, the distance to the $\mathrm{n^{th}}$ nearest neighbour. Defined in this way galaxies with closer neighbours, i.e. larger $\Sigma_n$, are located in denser environments. In this work we chose to utilize $\Sigma_5$, defined as
\begin{equation}
    \Sigma_5 = \frac{5}{\pi r^{2}_5}.
\end{equation}

The average medians of $\Sigma_5$, i.e. the critical values by which we split the sample into high and low, across all bins of distance are $\mathrm{\left<M(\Sigma_{5, all})\right>=4.01}$e-6, $\mathrm{\left<M(\Sigma_{5, R-type})\right>=3.66}$e-6, $\mathrm{\left<M(\Sigma_{5, S-type})\right>=5.58}$e-6, all in units of Mpc$^{-2}$~h.

It can be observed in the third row of Fig.~\ref{fig:eta_all} that we find no statistically significant difference in filtering the sample into high and low values of $\Sigma_5$. 
The high $\Sigma_5$ curve for the full void sample seems to be qualitatively similar to the analogous curve in Fig.~\ref{fig:eta_nosubsample}, while the low $\Sigma_5$ curve exhibits an even further dampening of the signal. This means that by selecting for high or low local density we are not affecting the detection of alignment signal, other than diluting it due to a lesser sample size. In other words, alignment seems to be independent of the local density of galaxies.

\subsection{Alignment signal dependence on void--centric velocities}
\label{sec:res_vel}

In this subsection we explore the relation between spin orientation systematics and galaxy void--centric 
velocities which could keep relics of preferred encounter direction and spin acquisition. 
We define radial and transverse velocities as

\begin{equation}
    v_\mathrm{rad} = \boldsymbol{v} \cdot \boldsymbol{\hat{r}}
\end{equation}

\begin{equation}
    v_\mathrm{tra} = \sqrt{ \boldsymbol{v}^2 - v_\mathrm{rad} }
\end{equation}

\noindent
respectively, where $\mathrm{\mathbf{v}}$ is the total velocity of the galaxy and $\mathrm{\mathbf{\hat{r}}}$ its void centric direction.

We find a significant difference between the low and high \textit{radial} velocity samples. The last row of Fig.~\ref{fig:eta_all} shows that a much higher perpendicular signal is found for galaxies with \textit{low} radial velocity. 
The difference in alignment signal between samples with low and high radial velocity is particularly strong in R--type voids. It is also noteworthy that for the all void types sample the signal persists above the 3$\sigma$ confidence region for every shell r > 1$\mathrm{R_v}$. 
On the other hand, we have also explored subsamples with low and high transverse velocity finding similar 
alignment signals in each case (not presented in the figure).
By comparison of the last row of Fig.~\ref{fig:eta_all} to the previous ones it can be observed 
that radial velocity appears as the galactic property that most strongly correlates with
the spin alignment signal, perpendicular to the void--centric radial direction.

\subsection{Spin alignment signals in combination of samples}
\label{sec:res_comb}

We have previously explored the alignment dependence on 5 galactic properties separately: spin norm, mass, $\Sigma_5$, transverse and radial velocity, finding the highest spin alignment signal for 
low void--centric radial velocity galaxies.
In this subsection we study spin alignments in all combination of subsamples considering high and low values of the parameters defining galactic properties. For a simpler cross--referencing we name these subsamples from one to 80 as ``Ss1'' (subsample one) to ``Ss80'' (subsample 80), with the entire galaxy sample with no selection regarding any galactic property being dubbed ``S0'' (sample zero).

For simplicity we also consider galaxies within a single shell
with a depth of 0.5$\mathrm{R_v}$, with inner and outer radii of 0.9 and 1.4$\mathrm{R_v}$ respectively, giving a
single value of the parameter $\zeta$ for each different void type. The results are shown in Tables \ref{table:1} and \ref{table:2}, where we highlight in bold fonts values with a large statistical significance, i.e. $|\zeta|>3$ (see Sec.~\ref{sec:etadef}).

Given that in the previous sections we find that low radial velocity is the the galactic property that most strongly correlates with perpendicular signal, we have divided the total set of results into two tables.
Table \ref{table:1} shows every possible combination of high and low galactic properties restricted to
high void--centric radial velocities, while the remaining set of subsamples with low radial velocity are shown in Table \ref{table:2}. 

%
Subsamples with alignment signals above the $3\sigma$ confidence level are highlighted in bold typeface across the Tables \ref{table:1} and \ref{table:2}.
These are S0 and subsamples with high mass, high spin, or low radial velocities. This result is consistent with the ones presented in previous subsections. 
The strongest spin alignment signals are found at approximately the 9$\sigma$ level for subsamples Ss55 and Ss57.
These two subsamples have in common high spin values and a low void--centric radial velocity selection (see Table~\ref{table:2}). 

The restriction of high or low $\Sigma_5$ galaxies dampen the previous signal--to--noise of the subsample; 
e.g., Ss9 and Ss18 with respect to S0, similarly as Ss17 and Ss26 with respect to Ss8. 
This further confirms our finding that local density, as measured with the $\Sigma_5$ parameter, is not directly correlated to spin alignment around voids.

Furthermore, to study the dependency of alignment signal with the other galactic properties, we can look at subsamples Ss1 (high spin), Ss8 (high mass), and Ss54 (low velocity).
We have $\zeta$=5.8\textpm1.0 for high spin and $\zeta$=5.6\textpm1.1 for high mass, for all void types, so none of these parameters correlates more strongly with alignment than the other. We conclude that 
only the selection of high mass and high spin galaxies has a strong incidence on
systematic spin alignments.
The strongest signal obtained for the three subsamples and for all void types, is
found for Ss54 at $\zeta$=6.8\textpm1.0, confirming our finding that a low void--centric
velocity is the greatest predictor of alignment amongst the parameters analysed.

With regards to void classification, we find more statistically significant signal values around R--type voids than around S--type voids. The highest value for R--type voids is found in Ss55 while the highest value for S--type voids belongs in Ss57. Both of these subsamples are low in radial velocity and can be seen in Table~\ref{table:2}. As seen in Sec.~\ref{sec:res_vel}, for a given distance from the void center, R--type voids inhabit a less dense environment than S--type voids. This means that we are detecting higher alignment signal in globally less dense environments.

\begin{figure}
    \includegraphics[width=1\columnwidth]{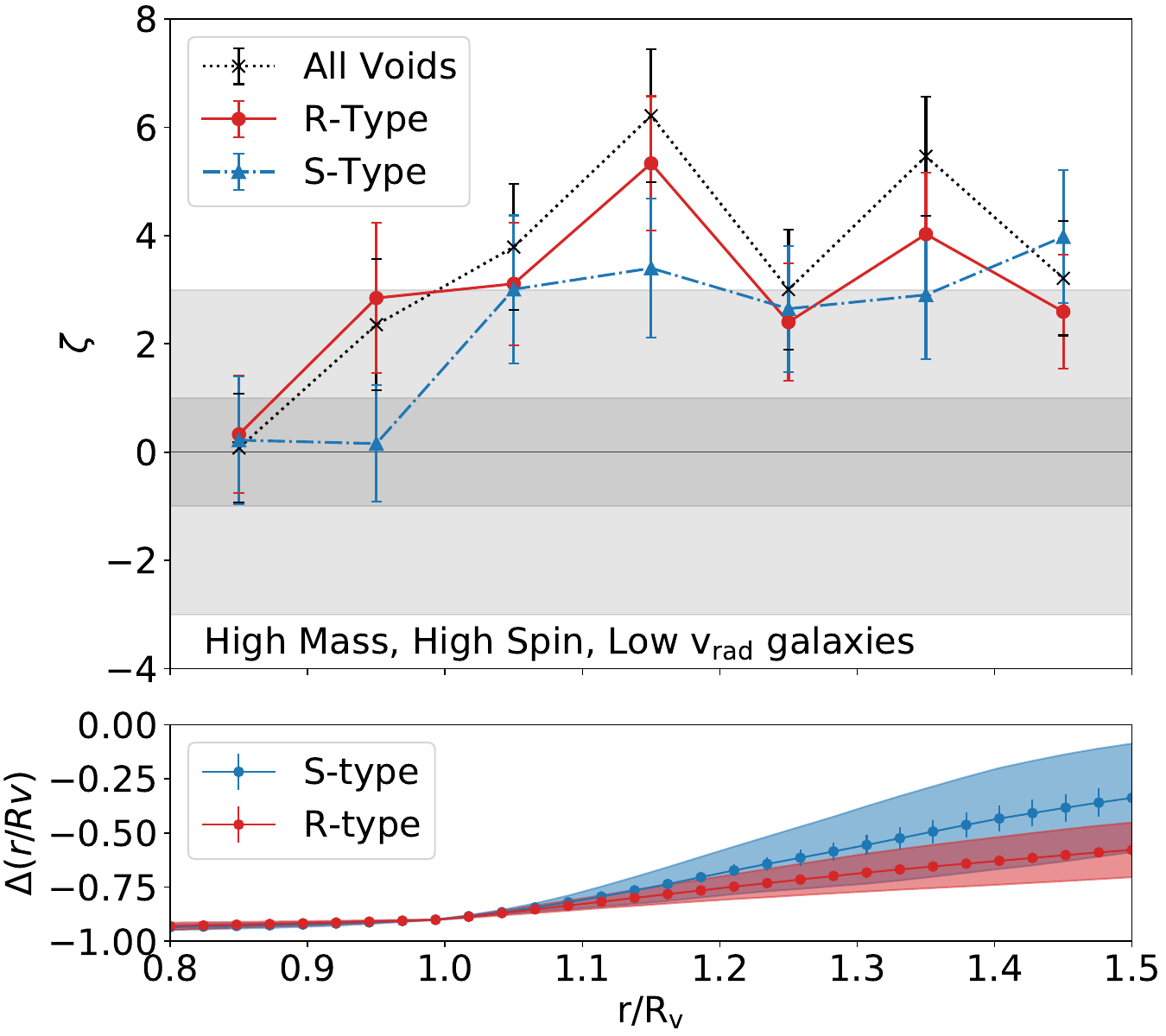}
\caption{From Table~\ref{table:1} and \ref{table:2} we
learn that the most aligned populations are massive galaxies with high spin and low radial velocity, particularly in R--type voids. In the top panel of this figure we plot the alignment signal $\zeta$ as a function of normalized distance to the center of the void. The black dotted line represents the signal of the stacking of all voids, while the solid red and dash-dot blue lines represent that of R-- and S--type voids respectively. We find a peak of perpendicular alignment signal of over 5$\sigma$ in the distance bin centered in 1.15$\mathrm{R_v}$ around R--type voids. The bottom panel shows that this is a very underdense region. In this panel we plot the cumulative density contrast in the same distance scale, and it is readily observable that, for R--type voids, the density contrast in this scale is $\mathrm{\Delta(1.15R_v)\simeq-0.75}$. The dots are the means of $\mathrm{\Delta(r)}$, while the error bars and shaded regions represent the errors of the mean and the standard deviation of the data respectively.}
\label{fig:bestsignal}
\end{figure}

We chose the subsample with the largest signal, Ss57 (high spin and mass, and low radial velocity), to plot its normalized alignment signal $\zeta$ as a function of distance to the center of the void in the top panel of Fig.~\ref{fig:bestsignal}. We find a large signal of over 5$\sigma$ for perpendicular alignment at around 1.15$\mathrm{R_v}$ in R--type voids. The bottom panel shows cumulative density contrast in the distances considered, and shows that the large signals are indeed found in underdense environments. 

The link between the alignment parameters analysed here, and more classic measurements such 
as the average cosine of the sample, is described in Dávila-Kurbán et al. (submitted). For Ss57, for example, we have $\mathrm{\zeta_{All Voids}}\simeq 9.0$ with N=23853 (see Table~\ref{table:2}); following the relation outlined in the aforementioned paper we can estimate that this value of $\mathrm{\zeta}$ corresponds to an average cosine value of $\mathrm{\langle cos(\theta) \rangle \simeq 0.43}$. 

%


\newcommand{\resaltar}[1]{\textbf{#1}}

\begin{table*}
\begin{tabular}{cllllcccrrr}
\hline
 Subsample Name & Spin & Mass & $\Sigma_5$ & $\mathrm{V_{rad}}$ & $\zeta_\mathrm{All\, Voids}$ & $\zeta_\mathrm{R-Voids}$ & $\zeta_\mathrm{S-Voids}$ & $\mathrm{N_{All}}$ & $\mathrm{N_R}$ & $\mathrm{N_S}$ \\
\hline
S0          & -     & -     & -           & -     & \resaltar{ 4.0 \textpm 1.0}  & \resaltar{ 3.6 \textpm 1.0}  &  1.8 \textpm 1.0             & 283573  & 217832  & 65741 \\
Ss1          & H     & -     & -           & -     & \resaltar{ 5.8 \textpm 1.0}  & \resaltar{ 5.0 \textpm 1.0}  &  3.0 \textpm 1.1             & 156192  & 120793  & 35399 \\
Ss2          & H     & L     & -           & -     & -0.1 \textpm 1.0             &  0.5 \textpm 1.0             & -1.1 \textpm 1.0             & 50217   & 39109   & 11108 \\
Ss3          & H     & H     & -           & -     & \resaltar{ 8.0 \textpm 1.1}  & \resaltar{ 6.6 \textpm 1.1}  & \resaltar{ 4.7 \textpm 1.1}  & 47707   & 36607   & 11100 \\
Ss4          & L     & -     & -           & -     & -0.4 \textpm 1.0             & -0.2 \textpm 1.0             & -0.5 \textpm 1.0             & 127381  & 97039   & 30342 \\
Ss5          & L     & L     & -           & -     & -0.4 \textpm 1.0             & -0.3 \textpm 1.0             & -0.1 \textpm 1.0             & 30580   & 23507   & 7073 \\
Ss6          & L     & H     & -           & -     &  0.0 \textpm 1.0             &  0.5 \textpm 1.0             & -0.8 \textpm 1.0             & 43598   & 32963   & 10635 \\
Ss7          & -     & L     & -           & -     & -0.4 \textpm 1.0             &  0.2 \textpm 1.0             & -1.0 \textpm 0.9             & 80797   & 62616   & 18181 \\
Ss8          & -     & H     & -           & -     & \resaltar{ 5.6 \textpm 1.1}  & \resaltar{ 5.0 \textpm 1.0}  &  2.6 \textpm 1.0             & 91305   & 69570   & 21735 \\
Ss9          & -     & -     & H           & -     & \resaltar{ 3.3 \textpm 1.1}  &  2.5 \textpm 1.0             &  2.0 \textpm 1.1             & 141785  & 108916  & 32870 \\
Ss10         & H     & -     & H           & -     & \resaltar{ 4.4 \textpm 1.0}  & \resaltar{ 3.5 \textpm 1.0}  & \resaltar{ 3.3 \textpm 1.1}  & 78096   & 60396   & 17699 \\
Ss11         & H     & L     & H           & -     & -0.3 \textpm 1.0             &  0.6 \textpm 1.0             & -1.9 \textpm 0.9             & 25108   & 19554   & 5554 \\
Ss12         & H     & H     & H           & -     & \resaltar{ 4.8 \textpm 1.1}  & \resaltar{ 3.1 \textpm 1.1}  & \resaltar{ 4.1 \textpm 1.1}  & 23852   & 18303   & 5550 \\
Ss13         & L     & -     & H           & -     & 0.0 \textpm 1.0             & 0.0 \textpm 1.0             & -0.1 \textpm 1.0             & 63690   & 48519   & 15171 \\
Ss14         & L     & L     & H           & -     & -0.4 \textpm 1.0             & -0.2 \textpm 1.0             &  0.1 \textpm 1.0             & 15290   & 11753   & 3536 \\
Ss15         & L     & H     & H           & -     &  0.2 \textpm 1.0             &  0.8 \textpm 1.0             & -0.1 \textpm 1.0             & 21799   & 16481   & 5317 \\
Ss16         & -     & L     & H           & -     & -0.4 \textpm 1.0             &  0.3 \textpm 1.1             & -1.4 \textpm 0.9             & 40398   & 31308   & 9090 \\
Ss17         & -     & H     & H           & -     & \resaltar{ 3.5 \textpm 1.1}  &  2.9 \textpm 1.0             &  2.8 \textpm 1.1             & 45652   & 34785   & 10867 \\
Ss18         & -     & -     & L           & -     &  2.3 \textpm 1.0             &  2.5 \textpm 1.1             &  0.6 \textpm 1.0             & 141786  & 108916  & 32870 \\
Ss19         & H     & -     & L           & -     & \resaltar{ 3.8 \textpm 1.0}  & \resaltar{ 3.6 \textpm 1.1}  &  0.9 \textpm 1.0             & 78096   & 60396   & 17699 \\
Ss20         & H     & L     & L           & -     &  0.1 \textpm 1.0             &  0.0 \textpm 1.0             &  0.4 \textpm 1.0             & 25108   & 19554   & 5554 \\
Ss21         & H     & H     & L           & -     & \resaltar{ 6.5 \textpm 1.1}  & \resaltar{ 6.2 \textpm 1.2}  &  2.6 \textpm 1.1             & 23853   & 18303   & 5550 \\
Ss22         & L     & -     & L           & -     & -0.4 \textpm 1.0             & -0.1 \textpm 1.0             & -0.5 \textpm 1.0             & 63690   & 48519   & 15171 \\
Ss23         & L     & L     & L           & -     & -0.2 \textpm 1.0             & -0.4 \textpm 1.0             & -0.2 \textpm 1.0             & 15290   & 11753   & 3536 \\
Ss24         & L     & H     & L           & -     & -0.2 \textpm 1.0             & -0.1 \textpm 1.0             & -0.9 \textpm 1.0             & 21799   & 16481   & 5317 \\
Ss25         & -     & L     & L           & -     & -0.1 \textpm 1.0             & -0.2 \textpm 1.0             &  0.1 \textpm 1.0             & 40398   & 31308   & 9090 \\
Ss26         & -     & H     & L           & -     & \resaltar{ 4.3 \textpm 1.0}  & \resaltar{ 4.2 \textpm 1.1}  &  1.0 \textpm 1.0             & 45652   & 34785   & 10867 \\
Ss27         & -     & -     & -           & H     & -1.1 \textpm 1.0             & -1.0 \textpm 1.0             & -0.3 \textpm 1.0             & 141786  & 108916  & 32870 \\
Ss28         & H     & -     & -           & H     & -0.7 \textpm 1.0             & -0.8 \textpm 1.0             & -0.2 \textpm 1.0             & 78096   & 60396   & 17699 \\
Ss29         & H     & L     & -           & H     & -1.9 \textpm 1.0             & -1.4 \textpm 1.0             & -1.5 \textpm 0.9             & 25108   & 19554   & 5554 \\
Ss30         & H     & H     & -           & H     &  2.5 \textpm 1.0             &  2.5 \textpm 1.1             &  1.1 \textpm 1.0             & 23853   & 18303   & 5550 \\
Ss31         & L     & -     & -           & H     & -0.8 \textpm 1.0             & -0.8 \textpm 1.0             & -0.5 \textpm 1.0             & 63690   & 48519   & 15171 \\
Ss32         & L     & L     & -           & H     &  0.4 \textpm 1.0             & 0.0 \textpm 1.0             &  1.2 \textpm 1.0             & 15290   & 11753   & 3536 \\
Ss33         & L     & H     & -           & H     & -0.8 \textpm 1.0             & -0.8 \textpm 1.0             & -0.4 \textpm 1.0             & 21799   & 16481   & 5317 \\
Ss34         & -     & L     & -           & H     & -1.3 \textpm 1.0             & -1.2 \textpm 1.0             & -0.6 \textpm 1.0             & 40398   & 31308   & 9090 \\
Ss35         & -     & H     & -           & H     &  1.2 \textpm 1.0             &  1.2 \textpm 1.0             &  0.3 \textpm 1.0             & 45652   & 34785   & 10867 \\
Ss36         & -     & -     & H           & H     & -0.4 \textpm 1.0             & -0.7 \textpm 1.0             &  0.2 \textpm 1.0             & 57201   & 45592   & 13017 \\
Ss37         & H     & -     & H           & H     & -0.7 \textpm 1.0             & -0.6 \textpm 1.0             &  0.5 \textpm 1.0             & 32361   & 25785   & 7196 \\
Ss38         & H     & L     & H           & H     & -1.2 \textpm 1.0             & -1.0 \textpm 1.0             & -2.0 \textpm 0.9             & 10359   & 8325    & 2248 \\
Ss39         & H     & H     & H           & H     &  1.0 \textpm 1.0             &  0.8 \textpm 1.0             &  1.9 \textpm 1.1             & 9972    & 7905    & 2308 \\
Ss40         & L     & -     & H           & H     & -0.5 \textpm 1.0             & -0.5 \textpm 1.0             & -0.2 \textpm 1.0             & 24819   & 19780   & 5822 \\
Ss41         & L     & L     & H           & H     & -0.6 \textpm 1.0             & -0.9 \textpm 1.0             & 0.0 \textpm 1.0             & 6056    & 4871    & 1357 \\
Ss42         & L     & H     & H           & H     &  0.1 \textpm 1.0             & 0.0 \textpm 1.0             & -0.6 \textpm 0.9             & 8521    & 6713    & 2059 \\
Ss43         & -     & L     & H           & H     & -1.4 \textpm 0.9             & -1.3 \textpm 1.0             & -1.4 \textpm 0.9             & 16425   & 13193   & 3599 \\
Ss44         & -     & H     & H           & H     &  1.0 \textpm 1.0             &  0.6 \textpm 1.0             &  1.0 \textpm 1.0             & 18507   & 14625   & 4343 \\
Ss45         & -     & -     & L           & H     & -1.0 \textpm 1.0             & -0.6 \textpm 1.0             & -0.5 \textpm 1.0             & 84583   & 63324   & 19852 \\
Ss46         & H     & -     & L           & H     & -0.3 \textpm 1.0             & -0.5 \textpm 1.0             & -0.7 \textpm 0.9             & 45735   & 34611   & 10502 \\
Ss47         & H     & L     & L           & H     & -1.4 \textpm 0.9             & -0.9 \textpm 1.0             & -0.2 \textpm 1.0             & 14749   & 11228   & 3306 \\
Ss48         & H     & H     & L           & H     &  2.5 \textpm 1.1             &  2.6 \textpm 1.1             & 0.0 \textpm 1.0             & 13879   & 10398   & 3242 \\
Ss49         & L     & -     & L           & H     & -0.6 \textpm 1.0             & -0.6 \textpm 1.0             & -0.4 \textpm 1.0             & 38870   & 28739   & 9349 \\
Ss50         & L     & L     & L           & H     &  1.0 \textpm 1.0             &  0.7 \textpm 1.0             &  1.6 \textpm 1.1             & 9234    & 6881    & 2179 \\
Ss51         & L     & H     & L           & H     & -1.1 \textpm 1.0             & -1.0 \textpm 1.0             & 0.0 \textpm 1.0             & 13278   & 9768    & 3258 \\
Ss52         & -     & L     & L           & H     & -0.5 \textpm 1.0             & -0.3 \textpm 1.0             &  0.3 \textpm 1.0             & 23973   & 18115   & 5491 \\
Ss53         & -     & H     & L           & H     &  0.9 \textpm 1.0             &  1.1 \textpm 1.0             & -0.3 \textpm 1.0             & 27144   & 20160   & 6523 \\

\hline
\end{tabular}
\caption{
This table shows the results for the measurement of $\zeta$ for all void types, as well as R-- and S--type voids separately for subsamples defined as all possible combinations of high and low Spin, Mass, and $\Sigma_5$, and radial velocity $\mathrm{V_{rad}}$, excluding samples with low $\mathrm{V_{rad}}$ which are shown in Table \ref{table:2}. The inner and outer radii of the shell considered is 0.9 and 1.4$\mathrm{R_v}$ respectively.}
\label{table:1}
\end{table*}

\begin{table*}
\begin{tabular}{cllllcccrrr}
\hline
 Subsample Name & Spin & Mass & $\Sigma_5$ & $\mathrm{V_{rad}}$ & $\zeta_\mathrm{All\, Voids}$ & $\zeta_\mathrm{R-Voids}$ & $\zeta_\mathrm{S-Voids}$ & $\mathrm{N_{All}}$ & $\mathrm{N_R}$ & $\mathrm{N_S}$ \\
\hline
Ss54         & -     & -     & -           & L     & \resaltar{ 6.8 \textpm 1.0}  & \resaltar{ 6.1 \textpm 1.0}  &  2.9 \textpm 1.0             & 141786  & 108916  & 32870 \\
Ss55         & H     & -     & -           & L     & \resaltar{ 9.1 \textpm 1.1}  & \resaltar{ 8.0 \textpm 1.1}  & \resaltar{ 4.5 \textpm 1.1}  & 78096   & 60396   & 17699 \\
Ss56         & H     & L     & -           & L     &  1.7 \textpm 1.0             &  2.1 \textpm 1.1             & 0.0 \textpm 1.0             & 25108   & 19554   & 5554 \\
Ss57         & H     & H     & -           & L     & \resaltar{ 9.0 \textpm 1.2}  & \resaltar{ 6.9 \textpm 1.2}  & \resaltar{ 5.7 \textpm 1.2}  & 23853   & 18303   & 5550 \\
Ss58         & L     & -     & -           & L     &  0.3 \textpm 1.0             &  0.7 \textpm 1.0             & -0.2 \textpm 1.0             & 63690   & 48519   & 15171 \\
Ss59         & L     & L     & -           & L     & -0.9 \textpm 1.0             & -0.4 \textpm 1.0             & -1.2 \textpm 1.0             & 15290   & 11753   & 3536 \\
Ss60         & L     & H     & -           & L     &  0.8 \textpm 1.0             &  1.6 \textpm 1.0             & -0.7 \textpm 1.0             & 21799   & 16481   & 5317 \\
Ss61         & -     & L     & -           & L     &  0.8 \textpm 1.0             &  1.3 \textpm 1.0             & -0.7 \textpm 1.0             & 40398   & 31308   & 9090 \\
Ss62         & -     & H     & -           & L     & \resaltar{ 6.7 \textpm 1.1}  & \resaltar{ 5.9 \textpm 1.1}  & \resaltar{ 3.5 \textpm 1.1}  & 45652   & 34785   & 10867 \\
Ss63         & -     & -     & H           & L     & \resaltar{ 4.6 \textpm 1.0}  & \resaltar{ 4.1 \textpm 1.1}  &  2.5 \textpm 1.1             & 84584   & 63324   & 19852 \\
Ss64         & H     & -     & H           & L     & \resaltar{ 6.5 \textpm 1.1}  & \resaltar{ 5.4 \textpm 1.1}  & \resaltar{ 3.9 \textpm 1.1}  & 45735   & 34610   & 10503 \\
Ss65         & H     & L     & H           & L     &  0.7 \textpm 1.0             &  1.8 \textpm 1.0             & -0.8 \textpm 1.0             & 14749   & 11229   & 3306 \\
Ss66         & H     & H     & H           & L     & \resaltar{ 5.6 \textpm 1.2}  & \resaltar{ 3.5 \textpm 1.1}  & \resaltar{ 3.8 \textpm 1.2}  & 13880   & 10398   & 3242 \\
Ss67         & L     & -     & H           & L     &  0.3 \textpm 1.0             &  0.3 \textpm 1.0             &  0.0 \textpm 1.0             & 38871   & 28739   & 9349 \\
Ss68         & L     & L     & H           & L     & -0.1 \textpm 1.0             &  0.6 \textpm 1.0             &  0.1 \textpm 1.0             & 9234    & 6882    & 2179 \\
Ss69         & L     & H     & H           & L     &  0.2 \textpm 1.0             &  1.1 \textpm 1.0             &  0.3 \textpm 1.0             & 13278   & 9767    & 3257 \\
Ss70         & -     & L     & H           & L     &  0.6 \textpm 1.0             &  1.7 \textpm 1.0             & -0.7 \textpm 1.0             & 23973   & 18115   & 5490 \\
Ss71         & -     & H     & H           & L     & \resaltar{ 3.9 \textpm 1.1}  & \resaltar{ 3.3 \textpm 1.1}  &  2.8 \textpm 1.1             & 27144   & 20160   & 6523 \\
Ss72         & -     & -     & L           & L     & \resaltar{ 5.0 \textpm 1.0}  & \resaltar{ 4.6 \textpm 1.1}  &  1.5 \textpm 1.0             & 57202   & 45592   & 13018 \\
Ss73         & H     & -     & L           & L     & \resaltar{ 6.5 \textpm 1.1}  & \resaltar{ 6.1 \textpm 1.1}  &  2.3 \textpm 1.0             & 32361   & 25785   & 7196 \\
Ss74         & H     & L     & L           & L     &  1.8 \textpm 1.1             &  1.2 \textpm 1.0             &  1.1 \textpm 1.0             & 10358   & 8325    & 2248 \\
Ss75         & H     & H     & L           & L     & \resaltar{ 7.4 \textpm 1.2}  & \resaltar{ 6.7 \textpm 1.2}  & \resaltar{ 4.5 \textpm 1.3}  & 9973    & 7904    & 2308 \\
Ss76         & L     & -     & L           & L     & 0.0 \textpm 1.0             &  0.5 \textpm 1.0             & -0.4 \textpm 1.0             & 24819   & 19779   & 5822 \\
Ss77         & L     & L     & L           & L     & -1.4 \textpm 0.9             & -1.3 \textpm 1.0             & -1.9 \textpm 0.9             & 6056    & 4871    & 1356 \\
Ss78         & L     & H     & L           & L     &  1.0 \textpm 1.1             &  1.0 \textpm 1.0             & -1.4 \textpm 0.9             & 8521    & 6713    & 2059 \\
Ss79         & -     & L     & L           & L     &  0.5 \textpm 1.0             &  0.0 \textpm 1.0             & -0.2 \textpm 1.0             & 16424   & 13193   & 3599 \\
Ss80         & -     & H     & L           & L     & \resaltar{ 6.0 \textpm 1.1}  & \resaltar{ 5.3 \textpm 1.2}  &  2.1 \textpm 1.1             & 18508   & 14625   & 4344 \\

\hline
\end{tabular}
\caption{This table shows the results for the measurement of $\zeta$ for all void types, as well as R-- and S--type voids separately for subsamples defined as all possible combinations of high and low Spin, Mass, and $\Sigma_5$ provided the radial velocity, $\mathrm{V_{rad}}$, be low. The inner and outer radii of the shell considered is 0.9 and 1.4$\mathrm{R_v}$ respectively. Overall, this set of subsamples with low radial velocities exhibits larger alignment signals.}
\label{table:2}
\end{table*}

\newcommand{\mybar}[3][3cm]{
\begin{tikzpicture}
\node[minimum width=#1] (top) {};
\node[fill=blue!10, minimum width={(#2/#3)*#1},
      below right=-10pt of top.south west] {#2};
\end{tikzpicture}
}



\section{Conclusions and discussion}
\label{sec:conclusions}

We have analyzed the orientations of galactic spins in subdense environments in the Illustris TNG300-1 simulation.
Our study shows a strong evidence that large galaxies in cosmic voids exhibit an excess of spins perpendicular to the void centric radial direction.
The statistics used to detect the alignment signal is robust and allows to explore its dependence on different regions in the parameter space, including the radial distance to the void center, spin magnitude, galaxy mass, local galaxy density, and the radial component of the velocity of galaxies relative to the cosmic void centre.

We find the highest alignment signal (at more than $9\sigma$ level) for massive galaxies with relevant rotation residing in void environments and with a low expansion velocity w.r.t. void centres. 
We stress that this sample of large galaxies with the highest rotation are the most reliable from the dynamical point of view of spin alignments. The fact that the strongest correlation is related to the void--centric expansion gives a hint that departures from the global dynamics of voids is a key ingredient to understand the origin of alignments.
Furthermore, we find that spin alignments are strongly dependent on the magnitude of the expansion velocity with respect to the void centre.
The fact that the most highly aligned spins are those in galaxies with a lower void--centric expansion velocity 
suggests that galaxies may gain an aligned spin as they lose linear momentum in their expansion away from the void center.
In this scenario with galaxy peculiar velocities having a strong contribution from void global expansion, 
the void--centric direction is privileged for galaxy encounters and accretion processes, a fact worth to study 
in future works.
On the other hand, the lack of dependence of the alignment results on $\Sigma_\mathrm{5}$
shows that the local galaxy density plays a minor role in the evolution of spin vectors.
The inclusion of void classification provides further hints on the origin of the effect. Our analysis show that R--type voids are those exhibiting the highest spin alignment effects. This is an indication that it is the void dynamics and its interaction with the evolving galaxies rather than the void surroundings that generates the systematic spin orientations.
  

In general, previous studies of spin alignments have been related to filaments or other over--dense structures or local environments. Here we detect alignment in under--dense regions as shown in Sec.~\ref{sec:galaxyvel}. 
Our finding of a preferential perpendicular orientation is consistent with the observational work of \citet{Trujillo2006}, which was later rebutted by similar works such as \citet{Slosar2009} and \citet{Varela2012}, pointing at a statistically small sample as the main reason for the discrepancy; however, this shortcoming is not present in our work. Furthermore, our findings are consistent with the predictions of TTT and observational studies of \citet{Lee2000, Lee2002} and \citet{Lee2007}, where, in the latter, the tidal tensor field is calculated and a preferential alignment for spins is found with its intermediate principal axis, which lies within the sheets (a proxy for our void surfaces). 
However, the alignment signal we find is particularly strong for galaxies that deviate from global void dynamics (low velocity seems to correlate with alignment), which can be due to encounters and would therefore be outside the scope of TTT.
On the other hand, we do not find this kind of alignment for massive galaxies as found by \citet{Codis2018, Kraljic2019}, most likely due to the vastly different environment densities these galaxies reside in.
Furthermore, we remark that, when taking into account massive galaxies, we find strong alignment signals only for those with high spin. We find no significant alignment signal for high--mass, \textit{low}--spin galaxies  (see Ss3 and Ss6 subsamples in Table~\ref{table:1}). Although this effect could be due to the more accurate determination of the spin axis in the case of high-spin galaxies, it could also hint at an important difference between galaxies with  high or low rotation-to-mass relation.
We notice, however, that a direct comparison between some of these previous works and the present paper is difficult to asses, since we have not performed a calculation of the tidal tensor field, and the void--centric direction can only be taken as a statistical proxy for the direction of the major principal axis. 

In future a work we will explore if these effects are redshift-dependent and whether the velocities and alignments correlate along the two-dimensional structure of void shells.

\section*{Acknowledgements}

This work was partially supported by the Consejo Nacional
de Investigaciones Cient\'ificas y T\'ecnicas (CONICET, Argentina), the Secretar\'ia de Ciencia y Tecnolog\'ia, Universidad Nacional de C\'ordoba, Argentina, and the Agencia Nacional de Promoci\'on de la Investigaci\'on, el Desarrollo Tecnol\'ogico y la Innovaci\'on, Ministerio de Ciencia, Tecnolog\'ia e Innovaci\'on, Argentina.
The IllustrisTNG project used in this work (TNG300) have been run on the HazelHen Cray XC40-system at the High Performance Computing Center Stuttgart as part of project GCS-ILLU of the Gauss centres for Supercomputing (GCS).
%
This research has made use of NASA’s Astrophysics Data System. Visualizations made use of python packages and inkscape software.

\section*{Data Availability}

The data underlying in this article are available on request to the corresponding author. 



\bibliographystyle{mnras}
\bibliography{references}


\bsp	
\label{lastpage}
\end{document}